\def\spacingset#1{\renewcommand{\baselinestretch}{#1}\small\normalsize}

\documentstyle{article}
\input mssymb.tex

\begin{document}
\newcommand{\nc}{\newcommand}
\nc{\rnc}{\renewcommand}
\nc{\nt}{\newtheorem}
\nc{\be}{\begin}
\nc{\erf}[1]{$\ (\ref{#1}) $}
\nc{\rf}[1]{$\ \ref{#1} $}
\nc{\lb}[1]{\mbox {$\label{#1}$} }
\nc{\hr}{\hrulefill}
\nc{\noi}{\noindent}

\nc{\eq}{\begin{equation}}
\nc{\en}{\end{equation}}
\nc{\eqa}{\begin{eqnarray}}
\nc{\ena}{\end{eqnarray}}

\nc{\ra}{\rightarrow}
\nc{\la}{\leftarrow}
\nc{\da}{\downarrow}
\nc{\ua}{\uparrow}
\nc{\Ra}{\Rightarrow}
\nc{\La}{\Leftarrow}
\nc{\Da}{\Downarrow}
\nc{\Ua}{\Uparrow}

\nc{\uda}{\updownarrow}
\nc{\Uda}{\Updownarrow}
\nc{\lra}{\longrightarrow}
\nc{\lla}{\longleftarrow}
\nc{\llra}{\longleftrightarrow}
\nc{\Lra}{\Longrightarrow}
\nc{\Lla}{\Longleftarrow}
\nc{\Llra}{\Longleftrightarrow}

\nc{\mt}{\mapsto}
\nc{\lmt}{\longmapsto}
\nc{\lt}{\leadsto}
\nc{\hla}{\hookleftarrow}
\nc{\hra}{\hookrightarrow}
\nc{\lgl}{\langle}
\nc{\rgl}{\rangle}

\nc{\stla}{\stackrel{d}{\la}}
\nc{\pard}{\partial \da}
\nc{\gdot}{\circle*{0.5}}


\rnc{\baselinestretch}{1.2}      
\nc{\bl}{\vspace{1ex}}           
\rnc{\theequation}{\arabic{section}.\arabic{equation}}  
\newcounter{xs}
\newcounter{ys}
\newcounter{os}
\nt{thm}{Theorem}[section]
\nt{dfn}[thm]{Definition}
\nt{pro}[thm]{Proposition}
\nt{cor}[thm]{Corollary}
\nt{con}[thm]{Conjecture}
\nt{lem}[thm]{Lemma}
\nt{rem}[thm]{Remark}

\nc{\Poincare}{\mbox {Poincar$\acute{\rm e}$} }

\nc{\bA}{\mbox{${\bf A}$\ }}
\nc{\bB}{\mbox{${\bf B}$\ }}
\nc{\bC}{\mbox{${\bf C}$\ }}
\nc{\bD}{\mbox{${\bf D}$\ }}
\nc{\bE}{\mbox{${\bf E}$\ }}
\nc{\bF}{\mbox{${\bf F}$\ }}
\nc{\bG}{\mbox{${\bf G}$\ }}
\nc{\bH}{\mbox{${\bf H}$\ }}
\nc{\bI}{\mbox{${\bf I}$\ }}
\nc{\bJ}{\mbox{${\bf J}$\ }}
\nc{\bK}{\mbox{${\bf K}$\ }}
\nc{\bL}{\mbox{${\bf L}$\ }}
\nc{\bM}{\mbox{${\bf M}$\ }}
\nc{\bN}{\mbox{${\bf N}$\ }}
\nc{\bO}{\mbox{${\bf O}$\ }}
\nc{\bP}{\mbox{${\bf P}$\ }}
\nc{\bQ}{\mbox{${\bf Q}$\ }}
\nc{\bR}{\mbox{${\bf R}$\ }}
\nc{\bS}{\mbox{${\bf S}$\ }}
\nc{\bT}{\mbox{${\bf T}$\ }}
\nc{\bU}{\mbox{${\bf U}$\ }}
\nc{\bV}{\mbox{${\bf V}$\ }}
\nc{\bW}{\mbox{${\bf W}$\ }}
\nc{\bX}{\mbox{${\bf X}$\ }}
\nc{\bY}{\mbox{${\bf Y}$\ }}
\nc{\bZ}{\mbox{${\bf Z}$\ }}

\nc{\cA}{\mbox{${\cal A}$\ }}
\nc{\cB}{\mbox{${\cal B}$\ }}
\nc{\cC}{\mbox{${\cal C}$\ }}
\nc{\cD}{\mbox{${\cal D}$\ }}
\nc{\cE}{\mbox{${\cal E}$\ }}
\nc{\cF}{\mbox{${\cal F}$\ }}
\nc{\cG}{\mbox{${\cal G}$\ }}
\nc{\cH}{\mbox{${\cal H}$\ }}
\nc{\cI}{\mbox{${\cal I}$\ }}
\nc{\cJ}{\mbox{${\cal J}$\ }}
\nc{\cK}{\mbox{${\cal K}$\ }}
\nc{\cL}{\mbox{${\cal L}$\ }}
\nc{\cM}{\mbox{${\cal M}$\ }}
\nc{\cN}{\mbox{${\cal N}$\ }}
\nc{\cO}{\mbox{${\cal O}$\ }}
\nc{\cP}{\mbox{${\cal P}$\ }}
\nc{\cQ}{\mbox{${\cal Q}$\ }}
\nc{\cR}{\mbox{${\cal R}$\ }}
\nc{\cS}{\mbox{${\cal S}$\ }}
\nc{\cT}{\mbox{${\cal T}$\ }}
\nc{\cU}{\mbox{${\cal U}$\ }}
\nc{\cV}{\mbox{${\cal V}$\ }}
\nc{\cW}{\mbox{${\cal W}$\ }}
\nc{\cX}{\mbox{${\cal X}$\ }}
\nc{\cY}{\mbox{${\cal Y}$\ }}
\nc{\cZ}{\mbox{${\cal Z}$\ }}

\nc{\rightcross}{\searrow \hspace{-1 em} \nearrow}
\nc{\leftcross}{\swarrow \hspace{-1 em} \nwarrow}
\nc{\upcross}{\nearrow \hspace{-1 em} \nwarrow}
\nc{\downcross}{\searrow \hspace{-1 em} \swarrow}
\nc{\prop}{| \hspace{-.5 em} \times}
\nc{\wh}{\widehat}
\nc{\wt}{\widetilde}
\nc{\nonum}{\nonumber}
\nc{\nnb}{\nonumber}

\nc{\half}{\mbox{$\frac{1}{2}$}}
\nc{\Cast}{\mbox{$C_{\frac{\infty}{2}+\ast}$}}
\nc{\Casth}{\mbox{$C_{\frac{\infty}{2}+\ast+\frac{1}{2}}$}}
\nc{\Casm}{\mbox{$C_{\frac{\infty}{2}-\ast}$}}
\nc{\Cr}{\mbox{$C_{\frac{\infty}{2}+r}$}}
\nc{\CN}{\mbox{$C_{\frac{\infty}{2}+N}$}}
\nc{\Cn}{\mbox{$C_{\frac{\infty}{2}+n}$}}
\nc{\Cmn}{\mbox{$C_{\frac{\infty}{2}-n}$}}
\nc{\Ci}{\mbox{$C_{\frac{\infty}{2}}$}}
\nc{\Hast}{\mbox{$H_{\frac{\infty}{2}+\ast}$}}
\nc{\Hasth}{\mbox{$H_{\frac{\infty}{2}+\ast+\frac{1}{2}}$}}
\nc{\Hasm}{\mbox{$H_{\frac{\infty}{2}-\ast}$}}
\nc{\Hr}{\mbox{$H_{\frac{\infty}{2}+r}$}}
\nc{\Hn}{\mbox{$H_{\frac{\infty}{2}+n}$}}
\nc{\Hmn}{\mbox{$H_{\frac{\infty}{2}-n}$}}
\nc{\HN}{\mbox{$H_{\frac{\infty}{2}+N}$}}
\nc{\HmN}{\mbox{$H_{\frac{\infty}{2}-N}$}}
\nc{\Hi}{\mbox{$H_{\frac{\infty}{2}}$}}
\nc{\Ogast}{\mbox{$\Omega_{\frac{\infty}{2}+\ast}$}}
\nc{\Ogi}{\mbox{$\Omega_{\frac{\infty}{2}}$}}
\nc{\Wedast}{\bigwedge_{\frac{\infty}{2}+\ast}}

\nc{\Fen}{\mbox{$F_{\xi,\eta}$}}
\nc{\Femn}{\mbox{$F_{\xi,-\eta}$}}
\nc{\Fp}{\mbox{$F_{0,p}$}}
\nc{\Fenp}{\mbox{$F_{\xi',\eta'}$}}
\nc{\Fuv}{\mbox{$F_{\mu,\nu}$}}
\nc{\Fuvp}{\mbox{$F_{\mu',\nu'}$}}
\nc{\cpq}{\mbox{$c_{p,q}$}}
\nc{\Drs}{\mbox{$\Delta_{r,s}$}}
\nc{\spq}{\mbox{$\sqrt{2pq}$}}
\nc{\Mcd}{\mbox{$M(c,\Delta)$}}
\nc{\Lcd}{\mbox{$L(c,\Delta)$}}
\nc{\Wxv}{\mbox{$W_{\chi,\nu}$}}
\nc{\vxv}{\mbox{$v_{\chi,\nu}$}}
\nc{\dd}{\mbox{$\widetilde{D}$}}

\nc{\diff}{\mbox{$\frac{d}{dz}$}}
\nc{\Lder}{\mbox{$L_{-1}$}}
\nc{\bone}{\mbox{${\bf 1}$}}
\nc{\px}{\mbox{${\partial_x}$}}
\nc{\py}{\mbox{${\partial_y}$}}

\setcounter{equation}{0}
\vspace{1in}
\begin{center}
{{\LARGE\bf
Some Classical and Quantum Algebras
 }}
\end{center}
\addtocounter{footnote}{0}
\vspace{1ex}
\begin{center}
Bong H. Lian and Gregg J. Zuckerman
\end{center}
\addtocounter{footnote}{0}
\footnotetext{G.J.Z. is supported by
NSF Grant DMS-9008459 and DOE Grant DE-FG0292ER25121.}

\vspace{1ex}
\begin{quote}
{\footnotesize
ABSTRACT.
We discuss the notion of a Batalin-Vilkovisky (BV) algebra and give
several classical examples from differential geometry and Lie theory.
We introduce the notion of a quantum operator algebra  (QOA) as a
generalization of  a  classical operator  algebra. In some examples,
we view a QOA  as a deformation of  a commutative algebra. We then
review the notion of a vertex operator algebra (VOA) and show that
a vertex operator algebra is  a QOA with some additional
structures.  Finally, we establish a connection between BV algebras
and VOAs.
}
\end{quote}
\addtocounter{footnote}{0}

\section{Introduction}

In reference \cite{LZ9},
the authors established a precise and general
connection between two types of algebras which are well known in
contemporary mathematical physics: Batalin-Vilkovisky algebras (BV algebras)
and vertex operator algebras (VOAs). BV algebras, although implicit
in modern mathematics, arose for the first time explicitly in the
context of BV quantization of classical field theories. VOAs arose
 for the first time in the parallel contexts of two dimensional
 conformal quantum field theory \cite{BPZ} and the mathematical
theory of monstrous moonshine \cite{Bor} \cite{FLM}.

The present paper may be regarded as a very brief mathematical introduction
 to both BV algebras and VOAs. We have attempted to relate both types
of algebras to objects and constructions in differential geometry,
supergeometry, Lie theory, commutative algebra, homological algebra
and operator algebras.

In section \ref{sec2a}, we point out that the algebra of differential
forms on a semi-Riemannian manifold is naturally a BV algebra.
Thus, BV algebras were implicit long ago in the absolute tensor
calculus and in general relativity. Likewise, the algebra of exterior
forms on a finite dimensional
 Lie algebra, which is endowed with a nondegenerate symmetric bilinear
form (or Hermitian form), is naturally a BV algebra.

As a pedagogical device in our discussion of VOAs, we introduce in
section \ref{sec2} the very simple and abstract notions of quantum operators
and quantum operator algebras (QOAs). Physicists have shied away
from such abstractions, but we have no such inhibitions. We have
found that many of the fundamental ideas and formulas in conformal
field theory can be very naturally explained in the setting of QOAs.

In order to connect QOAs to VOAs, we present in section \ref{sec4} a quick
discussion of three points of view on commutative algebras. We then
review the definition of a VOA in section \ref{sec5}, and discuss this
definition from three new points of view that parallel our previous
discussion in section \ref{sec4}. The main new result of the present
paper is Theorem \ref{translate}, which reformulates the notion of
a VOA (without a Virasoro quantum operator) as a particular and
remarkable type of QOA, which we call a creative QOA. We then
present two fundamental constructions of creative QOAs. The
second construction is a reformulation of some work of Igor
Frenkel and Yongchang Zhu \cite{FZ}. The approach of the present
paper was strongly influenced by the Frenkel-Zhu article.

In the final section of this paper (see Theorems \ref{5.3} and \ref{5.4}),
we restate the main theorems of \cite{LZ9} as theorems about conformal
QOAs (which are now equipped with a Virasoro quantum operator). In
brief, we employ the BRST precedure \cite{KO}\cite{Fe}
\cite{FGZ} to formulate
a cohomological construction of a BV algebra that starts from a choice
of conformal QOA with central charge twenty-six. Our main example
requires one ingredient: a simple Lie algebra over the complex
numbers. Our construction goes through without a hitch because of
the remarkable fact that there does not exist a simple Lie algebra
having dimension twenty-six.

Many other known constructions in string theory can in fact be
formulated in the language of QOAs and the BRST procedure
(see also \cite{Wi3}\cite{WZ}\cite{LZ3}\cite{LZ4}\cite{LZ5}).
The results on BV algebras in \cite{LZ9} have been applied in a recent
paper by Greg Moore \cite{M}.

The present
 paper grew out of a pair of lectures presented by the authors
in October 1993 to the University of North Carolina Mathematics
Department. We thank James Stasheff for the opportunity to give
these lectures. The authors are very pleased to dedicate this paper
to Professor Kostant on the occasion of his sixty-fifth birthday.

\section{Batalin-Vilkovisky Algebras}\lb{sec2a}

Let $A^*$ be a \bZ graded commutative associative algebra.
For every $a\in A$, let $l_a$ denote the linear map on $A$  given by
the left multiplication by $a$. Recall that a (graded) derivation $d$ on $A$ is
a homogeneous linear operator such that ${[d,l_a]} - l_{da}=0$ for all $a$.
A BV operator \cite{W2}\cite{Sch}\cite{GJ} $\Delta$ on $A^*$ is
a linear operator of degree -1 such that:\\
(i) $\Delta^2=0$;\\
(ii) ${[\Delta,l_a]} - l_{\Delta a}$ is a derivation on $A$ for all $a$, ie.
$\Delta$ is
a {\it second} order derivation.

A BV algebra is a pair $(A,\Delta)$ where $A$ is a graded commutative
algebra and $\Delta$ is a BV operator on $A$. The following is an elementary
but fundamental lemma:
\be{lem}\cite{GJ}\cite{Pen}
Given a BV algebra $(A,\Delta)$, define the BV bracket $\{,\}$ on $A$ by:
\[
(-1)^{|a|}\{a,b\} = {[\Delta,l_a]}b - l_{\Delta a}b.
\]
Then $\{,\}$ is a graded Lie bracket on $A$ of degree -1.
\end{lem}
By property (ii) above, it follows immediately that for every
$a\in A$, $\{a,-\}$ is a derivation on $A$. Thus a BV algebra is a sort of
odd Poisson algebra which, in mathematics, is also known as a Gerstenhaber
algebra \cite{Gers1}\cite{Gers2}.

\subsection{Some Classical Examples}

Let $M$ be an $n$-dimensional smooth manifold with a fixed volume form
$\omega$. Let
$C^*(M)$ be the deRham complex. Let $V^*(M)$ be the algebra of polyvector
fields -- ie. the exterior algebra on the smooth vector fields $Vect(M)$
over the the ring of smooth functions on $M$. There is a canonical
degree reversing linear
 isomorphism $i_\omega:V^*(M)\lra C^{n-*}(M)$ given by
the contraction with $\omega$: $v\mapsto i_\omega(v)$. Conjugating
the deRham differential $d$ by this isomorphism, we obtain a
square zero degree -1 operator $\Delta_\omega=i_\omega^{-1}d i_\omega$ on
$V^*(M)$.
In local coordinates $x^i$ for which the volume form is
$\omega=dx^1\wedge\cdots\wedge dx^n$,  we have
\eq\lb{bvop}
\Delta_\omega=\sum_i \frac{\partial}{\partial x^i} \frac{\partial}{\partial
{x^i}^*},
\en
where ${x^i}^*$ denotes the generator $\frac{\partial}{\partial x^i}$ in
$V^1(M)$.
It is evident that $\Delta_\omega$ is a second order derivation on $V^*(M)$,
hence
making $(V^*(M),\Delta_\omega)$ into a BV algebra \cite{Sch}.

Clearly the above construction generalizes to spaces in other categories:
algebraic, holomorphic etc.

The construction may be viewed slightly differently: we can regard
$V^*(M)$ as the commutative superalgebra of  functions on $\Pi T^*M$, the
cotangent bundle of $M$ with the fibers made into odd supervector spaces.
The BV bracket in $V^*(M)$ turns out to be equal to the
 odd Poisson bracket associated to the
canonical odd symplectic two-form on $\Pi T^*M$. The bracket is also
known as the Schouten bracket.

We should mention an important and well-known
application of the algebra $V^*(M)$.
Let $P$ be a bivector field on $M$. We can always construct a bracket
operation on the function algebra $C^\infty(M)$ by the formula
\eq
[f,g]_P = \iota_P(df\wedge dg)
\en
where $\iota_P$ denotes the contraction of $P$ against a two-form.
The question is: when does the new bracket $[,]$ give rise to a
Lie algebra structure on $C^\infty(M)$ (hence a Poisson algebra)?
\be{pro}
The bracket $[,]$ satisfies the Jacobi identity iff  the BV bracket
$\{P,P\}=0$.
\end{pro}

To discuss our second example, let's assume that the volume form $\omega$ above
comes from a metric $g$ on $M$. Let $d^*$ be the formal adjoint of
the deRham differential $d$, relative to the metric $g$. Note that $d^*$ is
a square zero degree -1 operator on $C^*(M)$.
Choose local coordinates so that the Riemannian volume form is
$\omega=dx^1\wedge\cdots\wedge dx^n$. Then once again
 $d^*$ has exactly the same form \erf{bvop} where
now ${x^i}^*$ denotes the generator $dx^i$ in the algebra $C^*(M)$.
Therefore $d^*$ is also a second order derivation making $(C^*(M), d^*)$
into a BV algebra which depends on $g$.

It turns out that this BV algebra is isomorphic to $(V^*(M), \Delta_\omega)$.
Locally this isomorphism is determined by
$dx^i\mapsto g^{ij}\frac{\partial}{\partial x^j}$.
The metric ensures that this is well-defined globally.

We now come to our third example, which arises in Lie theory. Let $\frak{g}$ be
any Lie algebra and $\bigwedge^*\frak{g}$ be its exterior algebra. Let $\delta$
be the Lie algebra homology differential on $\bigwedge^*\frak{g}$
\eq
\delta (X_1\wedge\cdots\wedge X_p) = \sum_{i<j}(-1)^{i+j}
{[X_i, X_j]}\wedge X_1\wedge
\cdots \hat{X}_i\cdots \hat{X}_j\cdots\wedge X_p.
\en
Then by direct computation
we verify that $\delta$ is a BV operator on the exterior algebra.

\section{Quantum Operator Algebras}\lb{sec2}

Let $V$ be a \bZ doubly graded vector space $V=\oplus V^n[m]$
such that $V[m]=0$ for all but finitely many negative $m$'s.
The degrees of a homogeneous element $v$ in $V^n[m]$ will
be denoted as $|v|=n$, $||v||=m$ respectively.
In physical applications, $|v|$ will be the fermion number of $v$.
In conformal field theory, $||v||$ will be the conformal dimension of $v$.

Let $z$ be
a formal variable with degrees $|z|=0$, $||z||=-1$. Then
it makes sense to speak of a {\it homogeneous} (biinfinite)
formal power series $a(z)=\sum_{n\in\bf Z}a(n)z^{-n-1}$
of degrees $|a(z)|$, $||a(z)||$ where
the coefficients $a(n)$ are linear maps in $V$ with degrees
$|a(n)|=|a(z)|$, $||a(n)||=-n-1+||a(z)||$. Note then that
the terms $a(n)z^{-n-1}$ indeed have the same degrees
$|a(z)|$, $||a(z)||$ for all $n$. We call a finite sum of such
homogeneous series $a(z)$ a {\it quantum operator} on $V$,
and we denote the linear space of quantum operator as
$QO(V)$.

Clearly it does not makes sense in general to multiply two
quantum operators $a(z)$, $b(z)$ pointwise. However the space $Lin(V)$
of graded linear maps in $V$ can be viewed as a subspace of $QO(V)$
where a linear map $A$ is regarded as the constant series
$\sum A\delta_{n,-1} z^{-n-1}$. The space $Lin(V)$ is
an associative algebra in a canonical way. Is there
a natural way to extend the product in $Lin(V)$ to all of $QO(V)$?
One such (nonassociative) extension is known as the
Wick product, defined as:
\eq
:a(z)b(z):=\sum_{n<0}a(n)z^{-n-1}b(z)+(-1)^{|a||b|}
b(z)\sum_{n\geq0}a(n)z^{-n-1}.
\en
It is evident that when restricted to $Lin(V)$, the Wick
product coincides with the natural product in $Lin(V)$.

Here we introduce a notation for iterated Wick products which we will
use later. Let
$a_1(z),...,a_n(z)$ be quantum operators. Their Wick product is
defined inductively as follows:
\eq
:a_1(z)\cdots\ a_n(z): \stackrel{def}{=} :a_1(z)(:a_2(z)\cdots a_n(z):):
\en

There is in fact a family of products (of which the Wick product
is one) which measure formally
the singularity of the formal product $a(z)b(w)$ as $z$
``approaches'' $w$.
\be{dfn}
For each integer $n$ we define
a product on $QO(V)$:
\eq
a(w)\circ_nb(w)=Res_z a(z)b(w)\iota_{z,w}(z-w)^n
-(-1)^{|a||b|}Res_zb(w)a(z)\iota_{w,z}(z-w)^n
\en
where
\eqa
\iota_{z,w}(z-w)^n
&=&\sum_{i\geq0}
\left(\be{array}{c}
n\\
i
\end{array}\right)
(-1)^iz^{n-i}w^i\nnb\\
\iota_{w,z}(z-w)^n &=&\sum_{i\geq0}
\left(\be{array}{c}
n\\
i
\end{array}\right)
(-1)^{n-i}z^iw^{n-i}.
\ena
\end{dfn}
To see that the above products are well-defined, take an element
$v$ in $V$ and consider first
\eq
Res_za(z)v\ \iota_{w,z}(z-w)^n =\sum_{i\geq0}
\left(\be{array}{c}
n\\
i
\end{array}\right)
(-1)^{n-i}w^{n-i}a(i)v.
\en
This is a finite sum because $a(i)v=0$ for all but finitely
many positive $i$. So
$Res_zb(w)a(z)v\ \iota_{w,z}(z-w)^n$ makes sense for any $v$ and
hence defines an element in $QO(V)$. Similarly for
$Res_za(z)b(z)v\ \iota_{z,w}(z-w)^n$. We note that
$a(z)\circ_{-1}b(z)$ is nothing but the Wick product $:a(z)b(z):$.
\be{pro}
For $a(z),b(z)$ in $QO(V)$, the following equality of
formal power series in two variables holds:
\eq\lb{ope}
a(z)b(w)=\sum_{n\geq0}a(w)\circ_n b(w) \iota_{z,w}(z-w)^{-n-1}
+:a(z)b(w):.
\en
\end{pro}
In this sense $:a(z)b(w):$ is the nonsingular part of
the {\it operator product expansion} \erf{ope} (see \cite{BPZ}), while
$a(w)\circ_n b(w) \iota_{z,w}(z-w)^{-n-1}$ is the polar part
of order $-n-1$.
We note that for $n<0$, we have
\eq\lb{deriv}
a(z)\circ_n b(z)=\frac{1}{(-n-1)!}:\partial^{-n-1}a(z) \ b(z):
\en
where $\partial=\frac{d}{dz}$. For $n\geq0$, we have
\eq
a(z)\circ_n b(z) = [(\sum_{m=0}^n
\left(\be{array}{c}
n\\
m
\end{array}\right)
a(m)(-z)^{n-m}),b(z)].
\en
\be{lem}\lb{comm}
Let $A$ be a  homogeneous linear operator on $V$. Then the commutator
$[A,-]$ is a graded derivation of each of the products $\circ_n$.
\end{lem}

Note that for $n=0$, we have $a(z)\circ_0 b(z) = [a(0),b(z)]$.
As a corollary, we have
\be{pro}
For any $a(z),b(z),c(z)$ in $QO(V)$ and $n$ integer, we have
\[
a(z)\circ_0(b(z)\circ_n c(z)) = {[ a(z)\circ_0 b(z)]} \circ_n c(z) +
(-1)^{|a||b|}
b(z)\circ_n {[a(z)\circ_0 c(z)]},
\]
ie. $a(z)\circ_0$ is a derivation of every product in $QO(V)$.
\end{pro}
The products $\circ_n$ will become important for describing
the algebraic and analytic structures of certain algebras
of quantum operators. Thus we introduce the following mathematical definitions:
\be{dfn}
 We say that $a(z),b(z)$ are mutually
local if $a(z)\circ_nb(z)=0$ for all but finitely many
positive $n$.
\end{dfn}

\be{dfn}\lb{2.5}
A graded subspace \cA \ of $QO(V)$ containing the identity operator
and closed with respect to
all the products $\circ_n$ is called a quantum operator algebra.
A QOA \cA is called local if its elements are pairwise mutually
local.
\end{dfn}
We observe that for any element $a(z)$ of a QOA, we have
$a(z)\circ_{-2}1 = \partial a(z)$. Thus a QOA is closed with respect
to formal differentiation.

\be{pro}\lb{2.6}
Suppose $O$ is a local QOA. Then for any $a(z),b(z)$ in $O$,
the matrix entries of $a(z)b(w)$ are rational.
\end{pro}
Proof: Let $v$ be in $V$ and $v^*$ in the restricted dual of $V$. By eqn
\erf{ope}, we have
\eq
\lgl v^*|a(z)b(w)|v\rgl = \sum_{n\geq0}\lgl v^*|a(w)\circ_n b(w)|v\rgl
(z-w)^{-n-1} +\lgl v^*|:a(z)b(w):|v\rgl.
\en
Each of the matrix entries, $\lgl v^*|a(w)\circ_n b(w)|v\rgl$ is a Laurent
polynomial
in $w$. If $a(z), b(z)$ are mutually local, then we have only finitely many
$(z-w)$-polar terms in the above expansion. Thus the singular part of
the operator product expansion contributes a rational function of
$z,w$ to the matrix entry. The matrix entry of the
Wick product is a Laurent polynomial. $\Box$

We note that none of the products $\circ_n$ is associative
in general. (We will return to this point later in an example.)
 However it clearly makes sense to speak of
the left, right or two sided ideals in a QOA and they are
defined in an obvious way.

We now return to a classical setting to motivate a construction
in the theory of QOAs.

\section{Commutative Algebras}\lb{sec4}

Let $A$ be a vector space with a distinguished nonzero element $\bone$.
We want to consider the set $C(A,\bone)$ consisting of commutative,
associative bilinear products $m$ on $A$ such that \bone\ is
the identity element for $m$. We want to relate this set
to some other sets naturally associated to the based space
$(A,\bone)$.

\be{dfn}\lb{3.1}
A translation map for $(A,\bone)$ is a linear map
$Y:A\lra Lin(A)$,
$a\mapsto Y(a)$, such that the following
axioms hold:

\noi(1)$Y(\bone)=id$\\
\noi(2)$Y(a)\bone=a$\\
\noi(3)$Y(a)Y(b)=Y(b)Y(a)=Y(Y(a)b)$.\\
Let $T(A,\bone)$ denote the set of all translation maps for
$(A,\bone)$.
\end{dfn}
A translation map $Y$ translates states (elements of $A$)
into operators (elements of $Lin(A)$). For each translation
map, we can define a product
$m_Y: A\otimes A\lra A$
\eq
m_Y(a,b)=Y(a)b.
\en
Then $m_Y$ is commutative and associative and \bone\
is the identity element. Moreover the map $Y$ is
an injective homomorphism of algebra $(A,m_Y,\bone)$
into the algebra $Lin(A)$.

Conversely if $m\in C(A,\bone)$,
we can define the map $Y_m: A\lra Lin(A)$
\eq
Y_m(a)b=m(a,b).
\en
Then it is clear that $Y_m$ is a translation map.
\be{pro}
The sets $C(A,\bone)$ and $T(A,\bone)$ are in bijective
correspondence via the map $m\mapsto Y_m$, and its
inverse $Y\mapsto m_Y$.
\end{pro}

\subsection{Creative Operator Algebras}

The above Proposition is of course trivial to prove, but it is
a stepping stone to the following idea. Let $Y$ be a
translation map. Let $O_Y$ denote the set of operators
$Y(A)$. We know that $O_Y$ is a commutative algebra
of linear operators on $A$ and that the map
$O_Y\lra A$, $Y(a)\lra Y(a)\bone$ is a linear isomorphism.

\be{dfn}
A creative operator algebra for the based space $(A,\bone)$
is a commutative subalgebra $O$ of $Lin(A)$ such that
the map $O\lra A$, $x\mapsto x\bone$ is a linear isomorphism.
We call this map the creative map associated to $O$. We write
$CO(V,\bone)$ for the set of creative operator algebras
for $(V,\bone)$.
\end{dfn}
It turns out that in the above definition,
it is enough to require that the creative map
is surjective.

If $O$ is a creative operator algebra,
let $Y_O:A\lra O\subset Lin(A)$ be
the inverse of the creative map $O\lra A$ associated with $O$.
Thus $Y_O(x\bone)=x$ for all $x\in O$, or equivalently
$Y_O(a)\bone=a$ for all $a\in A$. It is easy to check that
$Y_O$ is a translation map.
\be{pro}
The sets $T(A,\bone)$ and $CO(A,\bone)$ are in bijective
correspondence via the map $Y\mapsto O_Y$ and its inverse
map $O\mapsto Y_O$.
\end{pro}

To gain further insight into creative operator algebras
we state:
\be{pro}
If $O$ is in $CO(A,\bone)$, then\\
(i) $O$ is a maximal commutative subalgebra of $Lin(A)$;\\
\noi(ii) $O$ is a complementary subspace to the annihilator
$Ann(\bone)$ of \bone\ in $Lin(A)$.
\end{pro}

\section{Vertex Operator Algebras}\lb{sec5}

Recall that the formal variable $z$ is assigned degrees $|a|=0$, $||z||=-1$.
\be{dfn}\lb{3.3}\cite{Bor}\cite{FLM}
A VOA is a doubly graded vector space $V=\oplus V^k[l]$ with a distinguished
element $\bone$ with both degrees equal to zero, and a degree preserving
linear map
$Y(-,z):V\lra QO(V)$, satisfying
the following conditions:\\
(i) (Boundedness) For each $k$,
$V^k[l]=0$ for all but finitely many negative $l$;\\
(ii) (Unit) $Y(\bone,z)=id$;\\
(iii) (Nondegeneracy)
For any $v$ in $V$, $Y(v,z)\bone$ is a formal power series and
\[
lim_{z\ra0} Y(v,z)\bone = v.
\]
(iv) (Cauchy-Jacobi identity)
For any $v,v'$ in $V$, and integers $n,k,l$
\eqa
&&Res_{z-w} Y(Y(v,z-w)v',w)\ \iota_{w,z-w}z^n w^k (z-w)^l \nnb\\
&&=Res_z Y(v,z)Y(v',w)\ \iota_{z,w}z^n w^k (z-w)^l\nnb\\
&&-(-1)^{|v||{v'}|}Res_z Y(v',z')Y(v,z)\ \iota_{w,z}z^n w^k (z-w)^l
\ena
\end{dfn}
For now we will ignore
the so-called Virasoro structure on a VOA. (See below.)

We want to develop the theory of VOA's as a straightforward
but deep generalization of the theory of commutative algebras.
We have developed three points of view about commutative
algebras:\\
\noi 1. Bilinear product on a based space;\\
\noi 2. Translation map for a based space;\\
\noi 3. Creative operator algebra for a based space.

We have demonstrated the elementary equivalence of these
three points of view. Let's sketch informally what we know
about these three levels for VOA theory.

\noi 1. A VOA can be described in the
following terms:

\noi(i) A bigraded vector space $V=\oplus V^k[l]$ with
a distinguished element \bone of degrees $|\bone|=0=||\bone||$.
We will assume for simplicity that the second degree
is bounded from below.

\noi(ii) A family of bilinear products $m_n$, $n\in\bZ$, on $V$.
The degrees of $m_n$ are $|m_n|=0$, $||m_n||=-n-1$. Thus
for any pair of elements $a,b$ in $V$, $m_n(a,b)=0$ for
all but finitely many positive $n$. Only $m_{-1}$ is
of second degree zero.

\noi(iii) Axioms relating to \bone : for any $a$ in $V$,\\
a) for any nonnegative integer $n$,
$m_n(a,\bone)=0$\\
b) $m_{-1}(a,\bone)=a$\\
c) for any integer $n\neq-1$, $m_n(\bone,a)=0$\\
d) $m_{-1}(\bone,a)=a$.\\
Thus the operation $m_{-1}$ behaves most classically:
\bone\ is an identity element for $m_{-1}$. For the operation
$m_n$ with $n<-1$, \bone\ is only a right identity. For
the operation $m_n$ with $n>0$, \bone\ is a universal
left and right zero divisor.

\noi(iv) Cauchy-Jacobi axiom system: This is a complex system
of trilinear identities for the system of operations $m_n$.
Two kinds of terms appear: $m_n( a,m_p(b,c))$, $m_p( m_n(a,b),c)$
with various permutations of the three elements of $V$.

\be{rem}
(i) It would be very difficult at the present time to motivate the
Cauchy-Jacobi
axiom system in the purely classical setting of
systems of bilinear products. For one thing, it would be
difficult to give a simple and compelling example of an algebra satisfying all
of the above axioms. Moreover, the actual known examples of VOA's
did not arise in the language of this first level.

\noi(ii) It is already difficult to motivate algebras with more than two
bilinear products. We know that in some sense the operations $m_n$ in
a VOA fall into two classes: $n$  nonnegative, and $n$ negative.
We also know that in applications, the two operations, $m_0$ and $m_{-1}$
appear most often. However, the operation $m_1$ also occurs in our
work on BV operators (see below). Thus the BV operation in this language
is $m_1(b,v)$ where $b$ is a distinguished element.
\end{rem}

\noi 2. Associated to a VOA $V$ is the vertex map, $Y_m$, where $m$ now
stands for the sequence of operations $m_n$ in $V$:
\eq
Y_m(v,z)v' = \sum_n m_n(v,v') z^{-n-1}.
\en
To hide the direct reference to $m$,
we can introduce the mode operators $v(n)$ such that $||v(n)||=||v||-n-1$ if
$v$ is homogeneous, and such that
\eq
m_n(v,v')=v(n)v'.
\en
Finally we introduce the vertex operator
\eq
Y(v,z)=\sum_n v(n)z^{-n-1}
\en
for each $v$ in $V$. We can write the standard duality axioms for
a VOA in terms of the vertex operators above. We obtain axioms that look
like generalizations of the axioms for a translation map (Definition
\ref{3.1}).

\noi(i) $Y(\bone,z)=id$;\\
(ii) For any $v$ in $V$, $Y(v,z)\bone$ is a formal power series and
\[
lim_{z\ra0} Y(v,z)\bone = v.
\]
(iii) For any $v,v'$ in $V$ \cite{FLM}\cite{FHL},
\[
Y(v,z)Y(v',z')\sim (-1)^{|v||{v'}|}Y(v',z')Y(v,z) \sim Y(Y(v,z-z')v',z'),
\]
where the relation $\sim$ has to be defined carefully in terms of
the formal variable calculus. In particular we need to assume the
rationality of the matrix entries of the three formal expressions above.
(See Proposition \ref{2.6}.)

\noi{\bf Observation:} If all the vertex operators $Y(v,z)$ are constant
($z$-independent) operators in $V$, then $Y_m$ is exactly a translation
map in the sense of commutative algebra, and we can regard $V$ as a
graded commutative algebra with operation $m_{-1}$.

\noi 3. We want to take the third perspective: for each $v$
in $V$, $Y(v,z)$ is a ``quantum operator''. We want to ask several
questions about these quantum operators:\\

\noi(i) What sort of properties does $Y(v,z)$ have as a quantum operator on its
own?
In other words, what class of quantum operators arises from the study of
VOA's.\\
(ii) What properties does the space of
all vertex operators have as linear
subspace of the space of all quantum operators?

We recall that $QO=QO(V)$ denotes the space
of all quantum operators built from
$Lin(V)$. We write $QO_Y=QO(V,Y)$ as the space of all
quantum operators of the form $Y(v,z)$,
$v$  in $V$, for a fixed VOA structure
$Y$ on $V$.

Again we want to characterize in simple terms what sort of subspaces of
$QO$ can be of the form $QO_Y$ for some VOA structure $Y$ on $V$.
Eventually, we want to understand how to concretely construct examples
of VOA's by first constructing an appropriate subspace of $QO(V)$.
We need an appropriate generalization of the notion of a creative operator
algebra.

Recall that we have defined a sequence of bilinear operations $\circ_n$ in
$QO$.
It is a consequence of the Cauchy-Jacobi identity that the following holds:
\be{pro}
Suppose $(V,\bone,Y)$ is a VOA, and $v,v'$ are in $V$. Then for any
integer $n$,
\[
Y(v,z)\circ_n Y(v',z) = Y(v(n)v',z).
\]
\end{pro}
In particular, $QO_Y$ is closed with respect to {\it all} of the operations
$\circ_n$. Moreover for $v,v'$ in $V$,
\eq
Y(v,z)\circ_n Y(v',z)=0
\en
for all sufficiently large positive $n$.
This implies that $QO_Y$ is a {\it local}  quantum operator algebra
(See Definition \ref{2.5}). In particular, a vertex operator is
mutually local with itself. This is a highly nontrivial property for
a quantum operator.

Motivated by property 2(iii) of VOAs, we introduce the following:
\be{dfn}
A commutative quantum operator algebra $O$ is a local QOA such
that for any $a(z),b(z)$ in $O$,
\[
a(z)b(w)\sim (-1)^{|a||b|} b(w)a(z)
\]
in the sense that the matrix entries for both sides represent the same
rational function.
\end{dfn}

Let's  now introduce a distinguished element, \bone\ in the doubly
graded space $V$.
\be{dfn} (cf. Definition \ref{3.3})
A creative QOA for $(V,\bone)$ is a commutative QOA,  $O$, such that\\
(i) For every $a(z)$ in $O$, $a(z)\bone$ is a series with only nonnegative
powers of $z$.\\
(ii) The map $O\lra V$,
\[
a(z)\mapsto lim_{z\ra0} a(z)\bone
\]
is a linear isomorphism. This map is called the creative map associated with
$O$.
\end{dfn}
The reader should verify that every VOA, $(V,\bone,Y)$, defines a creative QOA,
$QO_Y$. We now give the converse, which is the main new result of this paper:
\be{thm}\lb{translate}
Let $(V,\bone)$ be a based space and let $O$ be a creative QOA for $(V,\bone)$.
Then $(V,\bone,Y)$ is a VOA where the vertex map $Y$ is given
by the inverse of the creative map associated to $O$.
\end{thm}
This theorem reformulates the Cauchy-Jacobi axiom in VOA theory in terms
of locality, commutativity and creativity in QOA theory. The reader
should compare our approach to that theorem of Frenkel-Lepowsky-Meurman
which reformulates the Cauchy-Jacobi axiom in terms of rationality,
associativity and commutativity \cite{FLM}.

\noi{\bf Question:} Given a commutative QOA, $O$, can we construct a
creative QOA?
\be{pro}
Let $O$ be a commutative QOA on the space $V$. Then for each nonzero
vector \bone\ of degrees $|\bone|=0, ||\bone||=0$, there is a canonical
subquotient $O_{\bf 1}$ of $O$ which is a creative QOA for some based space.
\end{pro}
To construct $O_{\bf 1}$, we proceed as follows: let $O'$ be the subspace
of quantum operators $a(z)$ in $O$ such that $a(z)\bone$ has only nonnegative
powers of $z$.
Let $V'$ be the image of the map $\chi:O' \lra V$, $a(z)\mapsto lim_{z\ra0}
a(z)\bone$.
We show that $O'$ is a closed with respective to all of the operations
$\circ_n$, hence is a commutative QOA on $V$. In fact by direct computation, we
have
for any $a(z),b(z)$ in $O'$, and for any integer $n$:
\eq\lb{cr}
lim_{z\ra0} a(z)\circ_n b(z)\bone = a(n)b(-1)\bone.
\en
This implies that the algebra $O'$ acts in $V'$ by restriction. This means that
$O'$ is now a commutative QOA for the based space $(V',\bone)$.

Now let $I$ be the two sided ideal generated by $ker \ \chi$ in $O'$. Consider
the subspace $\chi(I)$ in $V'$. If $a(z)$ is in $O'$ and $b(z)$ in $I$, then
$a(z)\circ_n b(z)$ is in $I$. It follows from eqn \erf{cr} that
$a(n)b(-1)\bone$
is in $\chi(I)$ for all $n$. Thus we have shown that the action of
$O'$ in $V'$ stabilizes the subspace $\chi(I)$. It is now clear that
$O'$ acts in $V'/\chi(I)$ in a natural way. Moreover, we have an induced
isomorphism $\chi: O_{\bf 1}=O'/I \lra V'/\chi(I)$. This map defines a creative
map
associated to the commutative QOA $O_{\bf 1}$ for the based space
$(V'/\chi(I),\bone)$.

\subsection{Examples}

Let $\cC$ be the Clifford algebra with the generators $b(n),c(n)$
($n\in\bZ$) and the relations \cite{FMS}
\be{eqnarray}
b(n)c(m)+c(m)b(n)&=&\delta_{n,-m-1}\nnb\\
b(n)b(m)+b(m)b(n)&=&0\nnb\\
c(n)c(m)+c(m)c(n)&=&0
\end{eqnarray}
Let $\lambda$ be a fixed integer.
The algebra \cC becomes \bZ-bigraded if we define the degrees
$|b(n)|=-|c(n)|=-1$, $||b(n)||=\lambda-n-1$, $||c(n)||=-\lambda-n$.
Let $\bigwedge^*$ be the graded irreducible
$\cC^*$-module with generator \bone\ and relations
\be{equation}
b(m)\bone=c(m)\bone=0, \ \ \ m\geq0
\end{equation}

Let $b(z),c(z)$ be the quantum operators
\be{eqnarray}
b(z)&=&\sum_mb(m)z^{-m-1}\nnb\\
c(z)&=&\sum_mc(m)z^{-m-1}
\end{eqnarray}
Let $O=O(b,c)$ be the smallest QOA containing $b(z),c(z)$.

\be{lem}
The QOA $O(b,c)$ is spanned by the following quantum operators
(see section \ref{sec2}):
\[
:\partial^{n_1}b(z)\cdots\partial^{n_i}b(z)\
\partial^{m_1}c(z)\cdots\partial^{m_j}c(z):
\]
with $n_1>...>n_i\geq0$, $m_1>...>m_j\geq0$.
\end{lem}
Let's sketch a proof. If $A(z)$ is in $O$, then so is
$A(z)\circ_{-2}1=\partial A(z)$ (see eqn \erf{deriv}).
Thus $O$ contains all the
derivatives of $b(z),c(z)$. Since $O$ is closed with respect
to the Wick product, the iterated
Wick products of the derivatives are also in $O$.
So it is enough to show that those iterated Wick products
are closed under every $\circ_n$, ie. if $A(z),B(z)$ are two
such products then $A(z)\circ_n B(z)$ is a sum of those products.
This can be shown by
induction and by computing
the operator product expansions in two ways:
$A(z)B(w)$ and $B(w)A(z)$.

By calculating the operator product expansion, one will
discover that in fact $O$ is {\it commutative as a QOA}.
Let's consider the map $\chi:O\lra \bigwedge$, $a(z)\mapsto
lim_{z\ra0}a(z)\bone$. Using the quantum operator given in the
above Lemma, we get
\eqa
&&lim_{z\ra0}:\partial^{n_1}b(z)\cdots\partial^{n_i}b(z)\
\partial^{m_1}c(z)\cdots\partial^{m_j}c(z):\bone
=n_1!\cdots n_i!m_1!\nnb\\
&&\cdots m_j!b(-n_1-1)\cdots b(-n_i-1)c(-m_1-1)\cdots c(-m_j-1)\bone.
\ena
Since the vectors on the right hand side form a basis for the space
$\bigwedge$, the map $\chi$ is necessarily an isomorphism. It follows
that $O$ is a creative QOA with the associated creative map $\chi$.
Thus by Theorem \ref{translate}, $(\bigwedge^*,\bone,\chi^{-1})$ is a VOA.

We remarked earlier that the products $\circ_n$ in a QOA are in general neither
associative nor commutative as bilinear operations. We can now give
a good example
to illustrate this. Let's rescale $c(z)$ by a parameter $\hbar$ (Planck's
constant) so that
we have
\eq
b(n)c(m)+c(m)b(n)=\hbar\delta_{n,-m-1}.
\en
It is clear that with this new relation the above construction still
goes through almost word for word, and we obtain a creative QOA $O_\hbar$.
Consider the following computations:
\eqa
:(:b(z)c(z):)b(z):-:b(z)(:c(z)b(z):)&=&\hbar\partial b(z)\nnb\\
:(:b(z)c(z):)b(z):-:b(z)(:b(z)c(z):)&=&\hbar\partial b(z).
\ena
The first eqn shows that the Wick product (ie. $\circ_{-1}$) is
nonassociative for
$\hbar\neq0$. The second eqn shows that it is noncommutative.
In fact the obstruction for associativity and commutativity is
of order $\hbar$. In this sense, $O_\hbar$ is a nonassociative deformation
of the commutative associative algebra $O_0$.

We now discuss our second example. Let $\frak{g}$ be a finite
dimensional simple
Lie algebra with the standard invariant bilinear form $B$. Let $\hat{\frak{g}}$
be the associated affine Kac-Moody Lie algebra whose bracket is
given by
\eq
[X(n),Y(m)]=[X,Y](n+m)+n\zeta B(X,Y)\delta_{n+m,0}
\en
where $\zeta$ is a nonzero element of the 1 dimensional center of
$\hat{\frak{g}}$. This Lie algebra can be given a \bZ grading
by $|X(n)|=0$, $||X(n)||=-n$ for all $X\in\frak{g}$.

Let $k$ be a complex parameter and $L({\frak{g}},k)$ be the graded
irreducible $\hat{\frak{g}}$-module generated by \bone, with
the relations:
\eq
X(n)\bone=0, \ \  n\geq0,X\in{\frak{g}},\ \ \zeta\bone=k\bone.
\en
For each $X$ in $\frak{g}$, define the quantum operator
\eq
X(z)=\sum_n X(n)z^{-n-1}.
\en
This has degrees $|X|=0$, $||X||=1$.
Let $O=O({\frak{g}},k)$ be the smallest QOA
in $QO(L({\frak{g}},k))$ such that $O$ contains
all the $X(z)$.
\be{pro}
$O({\frak{g}},k)$ is a creative QOA. If
$\{X_1,X_2,...\}$ is a basis for $\frak{g}$,
then $O({\frak{g}},k)$ is spanned by
\[
:\partial^{n_{11}}X_1(z)\partial^{n_{12}}X_1(z)\cdots
\partial^{n_{21}}X_2(z)\partial^{n_{22}}X_2(z)\cdots:
\]
where $n_{11}\geq n_{12}...$, $n_{21}\geq n_{22}...$,...
\end{pro}
The proof is similar to the first example above.

\be{cor}(\cite{FZ}, see remarks in introduction.)
$(L({\frak{g}},k),\bone,\chi^{-1})$ is a VOA, where $\chi$ is the
creative map associated to $O({\frak{g}},k)$.
\end{cor}

\section{Main Theorems}

We now discuss the main results, which draw from all the notions introduced
in this paper. We will also illustrate the results with some examples
which are built out of the previous ones.

Let $O$ be a creative QOA. A quantum operator $L(z)$ is called Virasoro if
for some constant $\kappa$ (called the central charge of $L$) \cite{BPZ},
\eqa
L(z)L(w) &=& \frac{\kappa}{2}(z-w)^{-4} + 2 L(w)(z-w)^{-2} +\partial L(w)
(z-w)^{-1} +:L(z)L(w):\nnb\\
L(z)a(w) &=& \cdots + ||a||a(w)(z-w)^{-2} + \partial a(w) (z-w)^{-1}+:L(z)a(w):
\ena
for all homogeneous $a(z)$ in $O$. Here ``$\cdots$'' denotes the higher order
polar terms.
\be{dfn}
A pair $(O,L)$ is called a conformal QOA if
$O$ is a creative QOA with a distinguished Virasoro quantum operator
 $L(z)$.
\end{dfn}

For example in the QOA $O(b,c)$ introduced above, for a fixed
$\lambda$, we have a Virasoro quantum operator \cite{FMS}
\eq
L^{bc}(z)=(\lambda-1):c(z)\partial b(z)+\lambda:\partial c(z)\ b(z):
\en
with central charge $\kappa=-12\lambda^2+12\lambda-2$.

In the second example $O({\frak{g}},k)$, if $k+h^\vee\neq0$ we define
(see \cite{KR})
\eq
L^{\frak{g}}(z)=\frac{1}{2(k+h^\vee)}\sum_i :X_i(z)X_i(z):
\en
where $h^\vee$ is the dual Coxeter number of $\frak{g}$ and
the $X_i$ is an orthonormal basis relative to the form $B$.
It is a fundamental fact the $L^{\frak{g}}(z)$ is a Virasoro quantum
operator.

\subsection{The BRST construction}

It is evident that if $(O,L)$, $(O',L')$ are conformal QOAs
on the respective based spaces $(V,\bone)$, $(V',\bone')$ with
central charges $\kappa,\kappa'$, then
$(O\otimes O',L+L')$ is a conformal QOA on $(V\otimes V',\bone\otimes\bone)$
with central charge $\kappa+\kappa'$. From now on we fix $\lambda=2$ which
means that $L^{bc}$ now has central charge -26. Let $(O,L)$
be any conformal QOA with central charge $\kappa$ and consider
\eq
C^*(O)= O^*(b,c)\otimes O.
\en
This QOA has a natural quantum operator called the BRST current \cite{FMS}
\cite{Fe}\cite{FGZ}:
\eq
J(z)=:c(z) (L(z)+\half L^{bc}(z)):.
\en
which has degrees $|J|=||J||=1$.
\be{lem}\cite{KO}\cite{Fe}\cite{FGZ}
Let $Q=Res_z J(z)$. Then $Q^2=0$ iff $\kappa=26$.
\end{lem}
Recall that for any element $a(z)$ in the QOA $C^*(O)$, we have
 $J(z)\circ_0 a(z)=[Q,a(z)]$. Thus
the graded commutator $[Q,-]$
is a derivation of the QOA $C^*(O)$. For $\kappa=26$, $[Q,-]$ is
a differential on the QOA $C^*(O)$ and
we have a cochain complex
\eq
[Q,-]:C^*(O)\lra C^{*+1}(O).
\en
It is called the BRST complex associated to $O$. Its cohomology will be denoted
as
$H^*(O)$. All the operations $\circ_n$ on $C^*(O)$
descend to the cohomology. However, all but one is trivial.
\be{thm}\lb{5.3}\cite{Wi3}\cite{WZ}\cite{LZ9}
The Wick product $\circ_{-1}$ induces a graded commutative product on
$H^*(O)$ with unit element represented by the identity operator. Moreover,
every cohomology class is represented by a quantum operator $a(z)$ with
$||a||=0$.
\end{thm}

Consider now the linear operator $\Delta: C^*(O)\lra C^{*-1}(O)$,
$a(z)\mapsto b(z)\circ_1 a(z)$.
\be{thm}\lb{5.4}\cite{LZ9}
The operator $\Delta$ descends to the cohomology $H^*(O)$. Morever,
it is a BV operator on the commutative algebra $H^*(O)$. Thus
$H^*(O)$ is naturally a BV algebra.
\end{thm}
The two main theorems above were originally proved in \cite{LZ9} in the
language of vertex operator algebras. It is Theorem \ref{translate}
which allows us to translate between the two versions. (For related
versions of Theorem \ref{5.4}, see \cite{GJ}\cite{SP}\cite{KSV}\cite{Hu}.)

\noi{\underline{Examples}:} Recall that $O({\frak{g}},k)$ is a conformal QOA
with a Virasoro quantum operator $L^{\frak{g}}(z)$. Its central charge
is $\kappa=\frac{k\ dim\ {\frak{g}}}{k+h^\vee}$.
Since the dimension of a simple Lie algebra $\frak{g}$ is never 26,
we can choose $k$ so that $\kappa=26$.
With this choice, we can consider the BRST cohomology
$H^*=H^*(O({\frak{g}},k))$. The answer is given as follows:
\eqa\lb{coho}
H^0&\cong& \bC\nnb\\
H^1&\cong& {\frak{g}}\nnb\\
H^2&\cong& {\frak{g}}\nnb\\
H^3&\cong& \bC.
\ena
Recall that given a Lie algebra, its exterior algebra is a BV algebra.
As a BV algebra, the structure of $H^*$ is determined by the following:
\be{pro}
There is a unique surjective homomorphism
of BV algebras $\bigwedge^*{\frak{g}}\lra
H^*(O({\frak{g}},k))$ such that under the above identification \erf{coho},
$X\mapsto X\in H^1$, for all $X$ in $\frak{g}$.
\end{pro}

\noi{\footnotesize DEPARTMENT OF MATHEMATICS, HARVARD UNIVERSITY
CAMBRIDGE, MA 02138. lian@math.harvard.edu}\\

\noi{\footnotesize DEPARTMENT OF MATHEMATICS, YALE UNIVERSITY
NEW HAVEN, CT 06520. gregg@math.yale.edu}


\begin{thebibliography}{99} \spacingset{1.2}

\bibitem{BPZ}
A. Belavin, A.M. Polyakov and A.A. Zamolodchikov,
``Infinite conformal symmetry in two dimensional quantum field theory'',
Nucl. Phys. B241 (1984) 333.

\bibitem{Bor}
R.E. Borcherds,
``Vertex operator algebras, Kac-Moody algebras and the Monster'',
Proc. Natl. Acad. Sci. USA. 83 (1986) 3068.

\bibitem{Fe}
B. Feigin,
``Semi-infinite homology for Virasoro and Kac-Moody algebras'',
Usp. Mat. Nauk. 39 (1984) 195-196.

\bibitem{FGZ}
I.B. Frenkel, H. Garland and G.J. Zuckerman,
``Semi-infinite cohomology and string theory'',
Proc. Nat. Acad. Sci. U.S.A. 83 (1986) 8442.

\bibitem{FLM}
I.B. Frenkel, J. Lepowsky and A. Meurman,
``Vertex Operator Algebras and the Monster'',
Academic Press, New York, 1988.

\bibitem{FHL}
I.B. Frenkel, Y. Huang and J. Lepowsky,
``On axiomatic approaches to vertex operator algebras and modules'',
Yale-Rutgers preprint, Dept. of Math., 1989.

\bibitem{FZ}
I.B. Frenkel and Y. Zhu,
``Vertex Operator Algebras associated to affine Lie algebras and the Virasoro
algebra'', Duke Math J. 66 (1992) 123.

\bibitem{FMS}
D. Friedan, E. Martinec and S. Shenker,
``Conformal invariance, supersymmetry and string theory'',
Nucl. Phys. B271 (1986) 93.

\bibitem{Gers1}
M. Gerstenhaber,
``The cohomology structure of an associative ring'',
Ann. of Math. 78, No.2 (1962) 267.

\bibitem{Gers2}
M. Gerstenhaber,
``On the deformation of rings and algebras'',
Ann. of Math. 79, No.1 (1964) 59.

\bibitem{GJ}
E. Getzler,
``Batalin-Vilkovisky algebras and two dimensional topological field theory'',
MIT preprint, hepth/9212043.

\bibitem{Hu}
Y-Z. Huang, ``Operadic formulation of topological vertex algebras and
Gerstenhaber or Batalin-Vilkovisky algebras'', Univ of Penn preprint,
hepth/9306021.

\bibitem{KR}
V. Kac and A. Raina,
{\it Highest weight representations of infinite dimensional Lie algebras},
Advanced Series in Math Physics, Vol. 2, World Scientific (1987).

\bibitem{KO}
M. Kato and K. Ogawa,
``Covariant quantization of string based on BRS invariance'',
Nucl. Phys. B212 (1983) 443.

\bibitem{KSV}
T. Kimura, J. Stasheff and A. Voronov,
``On operad structures of moduli spaces and string theory'',
UNC preprint, hepth/9307114.

\bibitem{LZ3}
B.H. Lian and G.J. Zuckerman,
``New selection rules and physical states in 2d gravity; conformal gauge'',
Phys. Lett B254, No.3,4 (1991) 417.

\bibitem{LZ4}
B.H. Lian and G.J. Zuckerman,
``2d gravity with c=1 matter'',Phys. Lett. B266 (1991) 21.

\bibitem{LZ5}
B.H. Lian and G.J. Zuckerman,
``Semi-infinite homology and 2D gravity (I)'',
Commun. Math. Phys. 145 (1992) 561.

\bibitem{LZ9}
B.H. Lian and G.J. Zuckerman,
``New perspectives on the BRST-algebraic structure in string theory'',
hepth/9211072, Commun. Math. Phys. 154 (1993) 613.

\bibitem{M}
G. Moore,
``Finite in All Directions'',
Yale preprint, hepth/9305139.

\bibitem{Pen}
M. Penkava,
``A note on BV algebras'',
UC Davis preprint November 92.

\bibitem{SP}
M. Penkava and A. Schwarz,
``On some algebraic structures arising in string theory'', UC Davis preprint,
hepth/9212072.

\bibitem{Sch}
A. Schwarz,
``Geometry of Batalin-Vilkovisky quantization'',
UC Davis preprint, hepth/9205088.

%
\bibitem{Wi3}
E. Witten,
``Ground ring of the two dimensional string theory'',
hepth/9108004, Nucl. Phys. B373 (1992) 187.

\bibitem{W2}
E. Witten,
``A note on the anti-bracket formalism'',
preprint IASSNS-HEP-90/9.

\bibitem{WZ}
E. Witten and B. Zwiebach,
``Algebraic structures and differential geometry in 2d string theory'',
hepth/9201056, Nucl. Phys. B377 (1992) 55.

\end{thebibliography}
\end{document}